# A STUDY OF THE GRAVITATIONAL WAVE FORM FROM PULSARS

**S. R. VALLURI**, *Department of Physics & Astronomy and Department of Applied Mathematics, University of Western Ontario, London, Ontario, N6A 5B7, Canada, Email:* valluri@uwo.ca, **J. J. DROZD**, *Department of Applied Mathematics, University of Western Ontario, London, Ontario, Email:* jdrozd1@uwo.ca , **F. A. CHISHTIE**, *Department of Applied Mathematics, University of Western Ontario, Email:* fachisht@uwo.ca, **R. G. BIGGS**, *Department of Mathematics, University of Western Ontario,* **M. DAVISON**, *Department of Applied Mathematics, University of Western Ontario, Email:* mdavison@uwo.ca, **SANJEEV V. DHURANDHAR**, *Inter-University Centre for Astronomy and Astrophysics, (IUCAA), Post Bag 4, Ganeshkhind, Pune 411 007, India, Email:* sdh@iucaa.ernet.in, **B. S. SATHYAPRAKASH**, *Department of Physics and Astronomy, University of Cardiff, UK, Email:* b.s.sathyaprakash@astro.cf.ac.uk.

We present analytical and numerical studies of the Fourier transform (FT) of the gravitational wave (GW) signal from a pulsar, taking into account the rotation and orbital motion of the Earth. We also briefly discuss the Zak-Gelfand Integral Transform. The Zak-Gelfand Integral Transform that arises in our analytic approach has also been useful for Schrodinger operators in periodic potentials in condensed matter physics (Bloch wave functions).

## 1 INTRODUCTION

The direct detection of gravitational radiation (GR) from astrophysical sources is one of the most important outstanding problems in experimental gravitation today. The construction of large laser interferometric gravitational wave detectors like the LIGO[1], VIRGO[2], LISA, TAMA 300, GITO 6000 and AIGO is opening a new window for the study of a vast and rich variety of nonlinear curvature phenomena. The network of gravitational wave detectors can confirm that GW exist and by monitoring gravitational wave forms give important information on their amplitudes, frequencies and other important physical parameters.

A prototype of continuous astrophysical sources is a pulsar. A variety of instabilities cause deformations from spherical symmetry giving rise to GWs. The amplitudes of GR from these pulsars are probably very weak ($\leq 10^{-26} - 10^{-28}$, for galactic pulsars). The GR signal will be buried deep within the noise of the detector system. The detection of a GR signal warrants the urgent need of careful data analysis with development of *analytical methods* and *problem oriented algorithms.*

In sections 2, 3 and 4 we outline the approach leading to the FT of the GW signal. The frequency modulation (FM), Doppler shift due to rotation and orbital motion of the Earth in the Solar System Barycentre (SSB) frame, its effect on the total phase of the received GW signal and the Fourier transform (FT) of the GW signal have previously been described[2].

Sections 5 and 6 present discussion and conclusions.

## 2 Methodology

Typical values of the gravitational wave amplitude *h* for the Crab and Vela pulsars are $\sim 10^{-25}$ and $\sim 10^{-24}$ respectively. This amplitude is several orders of magnitude below LIGO's expected sensitivity of $\sim 10^{-23}$). Since the LIGO would make continuous observations over a time scale of a few months or more, a significant enhancement to the signal-to-noise ratio (SNR) is expected by integrating the data over a long time interval.

The total response of the detector is a function of the source position, the detector orientation, the orientation of the spin axis of the Earth and the orientation of the orbital plane. Since the pulsar signal is weak, long interaction times $\approx 10^7$ secs will be needed to extract the signal from the noise. Since the detector moves along with the Earth in this time, the frequency of the wave4 emitted by the source is Doppler shifted. Also since the detector has an anisotropic response, the signal recorded by the detector is both frequency and amplitude modulated. We discuss now the important role of frequency modulation in the context of signal detection.

## 3  Study of Frequency Modulated Pulsar Signal

Frequency modulation arises due to translatory motion of the detector acquired from the motion of the Earth. We consider only two motions of the Earth: its rotation about the spin axis and the orbital motion about the Sun, so the response is doubly frequency modulated with one period corresponding to a day and the other period corresponding to a year. The FM smears out a monochromatic signal into a small bandwidth around the signal frequency of the monochromatic waves. It also redistributes the power in a small bandwidth. The study of FM due to rotation of the Earth about its spin axis for a one day observation period shows that the Doppler spread in the angular bandwidth for 1kHz signal will be 0.029Hz. The Doppler spread in the angular bandwidth due to orbital motion for an observation period of one day will be $1.7 \times 10^{-3}$Hz [2].

Since any observation is likely to last longer than a day, *it is important to incorporate this effect in the data analysis algorithms.*

In order to study frequency modulation of a monochromatic plane wave, one must calculate the Doppler shift due to rotational and orbital motion of the Earth in the SSB frame. For this, we need to know the relative velocity between the source and detector. The Euler angles ($\theta$, $\phi$) give the direction of the incoming wave in the SSB frame. We characterize the motion of the Earth (and detector) simply: (a) We assume the orbit of the Earth to be circular. (b) We neglect the effect of the Moon and the perturbation effects of Jupiter on the Earth's orbit.

The phase $\phi(t)$ of the Doppler shifted received signal for a single direction sky search ($\theta$, $\phi$) is given by,

$$\phi(t) = 2\pi \int_{t_0}^{t} f_{rec}(t')dt'$$

$$= 2\pi f_0 \int_{t_0}^{t} \left(1 + \frac{\vec{v} \bullet \vec{n}}{c}(t')\right) dt' \qquad (1)$$

$$= 2\pi f_0 \begin{bmatrix} t - t_0 + \left\{\frac{A}{C}\sin\vartheta\cos\phi' + \frac{R}{C}\sin\alpha\{\sin\vartheta(\sin\beta'\cos\varepsilon\sin\phi + \cos\phi\cos\beta') + \sin\beta'\sin\varepsilon\cos\vartheta\}\right\} \\ - \left\{\frac{A}{C}\sin\vartheta\cos\phi_0' + \frac{R}{C}\sin\alpha\{\sin\theta(\sin\beta_0'\cos\varepsilon\sin\phi + \cos\phi\cos\beta_0') + \sin\beta_0'\sin\varepsilon\cos\vartheta\}\right\} \end{bmatrix}$$

where $\phi' = \omega_{orb}t - \phi$, $\beta' = \beta_0 + \omega_{rot}t$, $\phi_0' = \omega_{orb}t_0 - \phi$, $\beta_0' = \beta_0 + \omega_{rot}t_0$, $\beta_0$ is the initial azimuthal angle of the detector at the observation time $t_0$. where $A$ is the distance from the centre of the SSB frame to the centre of the Earth, $R$ is the radius of the Earth $(R_\varepsilon)$, and $\vec{n}$ the unit vector in the direction of source, $\vec{n} = (\sin\vartheta\cos\phi, \sin\vartheta\sin\phi, \cos\vartheta)$. Here we have assumed that at time $t = 0$, the longitudinal angle $\beta = 0$.

It can be seen from eq (1) that the Doppler corrections to the phase of received pulsar signal depends on the direction of the source in the sky. The Earth's orbital motion is now included in this analysis which is a development of our previous analysis which considered only rotational motion.

## 4  Fourier Transform Analysis of the FM signal due to the Rotational and Orbital Motion of the Earth

We analyze the Fourier transform (FT) of the frequency modulated signal and study the extent to which the peak of the FT is smudged and to how much the FT *spreads* in the frequency space. This type of study would be useful from the point of view of data analysis and for applying such schemes as *stepping around the sky* method[3] which relies on the FT.

We let $x = \frac{2\pi f_0 R}{c}$ and $t_0 = 0$. Here $x$ plays the role of modulation index similar to $K$ in the theory of signal modulation. The modulation index depends on the frequency of the pulsar signal, the output of amplitude unity is given as follows,

$$h(t) = \cos(\phi(t)) \qquad (2)$$

We now consider the $h(t)$ to be given on a finite time interval $[0,T]$ which would be assumed to be the observation period. In our previous analysis we have assumed $T$ to be one day. Now $T= 365$ days. The Fourier transform of the signal $h(t)$ is given by,

$$\tilde{h}(f) = \int_0^T h(t) e^{-i 2\pi f t} dt \quad (3)$$

It is convenient to use a time coordinate $\xi = \omega_{rot} t / 2$ which for a period of a day is of the order of unity, i.e. when $T = 1\text{day} = 84600$ secs, the $\xi_T = \omega_{rot} T = \pi$, $(\omega_{rot} = \omega_r \text{ henceforth})$.

An exact closed form expression for the Fourier transform of the frequency modulated GW signal is obtained by the analytical approach. The plane wave expansion in spherical harmonics is used.[4]

The Doppler shift is given by the expression $f = f_0 \left[ 1 + \frac{\hat{n} \cdot \dot{\bar{r}}}{c} \bigg|_{t_d} \right]$ (4)

where
$\hat{n} = (Sin\theta Cos\phi, Sin\theta Sin\phi, Cos\theta)$, the unit vector in the source direction,
$t_0 = 0 \text{ results in } \beta = 0$,
$\frac{v \cdot n}{c}(t) = \frac{\dot{r}_{tot}(t) \cdot \hat{n}}{c}$ is the Total Doppler Shift at time $t$ due to rotation and the orbital motion of the earth (SSB frame),

$$\bar{r}_{tot}(t) - \bar{r}_{tot}(0) = \left[ \begin{pmatrix} A(Cos(\omega_{orb}t)-1) \\ A(Sin(\omega_{orb}t)) \\ 0 \end{pmatrix} + \begin{pmatrix} R Sin\alpha (Cos(\omega_{rot}t)-1) \\ R Sin\alpha Sin(\omega_{rot}t) Cos\varepsilon \\ R Sin\alpha Sin(\omega_{rot}t) Sin\varepsilon \end{pmatrix} \right] = \bar{R}_{orb} + \bar{R}_{rot} \quad (5)$$

We need to consider the expression:

$$e^{i\phi_{phase}(t)} = e^{i 2\pi f_0 \left\{ t + \hat{n} \cdot \left( \frac{\bar{R}_{orb} + \bar{R}_{rot}}{c} \right) \right\}}$$

$$= \exp\{i 2\pi f_0 t\} \exp\left\{ i 2\pi f_0 \left( \hat{n} \cdot \frac{\bar{R}_{rot}}{c} \right) \right\} \exp\left\{ i 2\pi f_0 \left( \hat{n} \cdot \frac{\bar{R}_{orb}}{c} \right) \right\} \quad (6)$$

The second and third of these exponential expressions consider the purely rotational and orbital effect terms.

The plane wave expansion for $\exp\left\{ i 2\pi f_0 \frac{\hat{n}}{c} \cdot \bar{R}_{rot} \right\}$ is of the form:

$$4\pi \sum_{\ell m} i^\ell Y_{\ell m}(\theta, \phi) Y_{\ell m}^*\left( \alpha, \frac{\omega_r t}{2} \right) j_\ell\left( k \, Sin\left( \frac{\omega_r t}{2} \right) \right), \quad -\ell \leq m \leq \ell, \ 0 \leq \ell \leq \infty \quad (7)$$

Here,
$k = \frac{4\pi f_0 R_\varepsilon Sin\alpha}{c}$, $\alpha$ is the colatitude of the detector, $j_\ell$ is the spherical Bessel function and $Y_{\ell m}$ are the Spherical Harmonics.

Now consider the orbital part. Its expansion gives:

$$\exp\left\{ i \frac{2\pi f_0}{c} \bar{R}_{orb}(t) \cdot \hat{n} \right\} = \exp\left[ i \frac{2\pi f_0 A}{c} Sin\theta \cdot \{Cos(\omega_{orb}t)Cos\phi + Sin(\omega_{orb}t)Sin\phi - Cos\phi\} \right]$$

$$= e^{-i \frac{2\pi f_0 A}{c} Sin\theta Cos\phi} \cdot e^{i \frac{2\pi f_0 A}{c} Cos(\omega_{orb}t - \phi) Sin\theta} \quad (8)$$

$$= \underbrace{e^{-i \frac{2\pi f_0 A}{c} Sin\theta Cos\phi}}_{\substack{\text{Independent of } t, \\ \text{depends on } f_0, A, \text{ and } \phi}} \cdot \sum_{n=-\infty}^{\infty} i^n \underbrace{e^{in\omega_{orb}t}}_{\substack{\text{will be combined} \\ \text{with } e^{i 2\pi f_0 t} e^{i \frac{m}{2} \omega_r t}}} e^{-in\phi} \underbrace{J_n\left( \frac{2\pi f_0 A}{c} \cdot Sin\theta \right)}_{\text{Independent of } t}$$

The result and expression for the Fourier transform of the Signal $S_{n\ell m}(f_0, \alpha, t)$ can be written as:

$$\sum_{n,\ell,m} S_{n\ell m}(\omega_0,\omega_r,\omega_{orb},\alpha,t,A,k,\theta,\phi) = \begin{bmatrix} \sum_{n\ell m} 4\pi i^\ell Y_{\ell m}(\theta,\phi) N_{\ell m} P_\ell^m(\cos\alpha) \sqrt{\frac{\pi}{2}} e^{-i\frac{2\pi A f_0}{c}\sin\theta\cos\phi} \\ \bullet \int_0^T e^{-i2\pi(f-f_0)t} \frac{J_{\ell+\frac{1}{2}}\left(k\sin\left(\frac{\omega_r t}{2}\right)\right)}{\sqrt{k\sin\left(\frac{\omega_r t}{2}\right)}} e^{-im\frac{\omega_r}{2}t - in\omega_{orb}t} dt \\ \bullet i^n e^{-in\phi} J_r\left(\frac{2\pi f_0 A \sin\theta}{c}\right) \end{bmatrix} \quad (9)$$

where,

$2\pi(f - f_0) = \omega - \omega_0$ and $N_{\ell m} = \sqrt{\frac{(2\ell+1)(\ell-|m|)!}{4\pi(\ell+|m|)!}}$ is the normalization constant for the Legendre polynomials $P_\ell^m$,

and $J$ is the cylindrical Bessel function

We now let $T = 365$ days. For $T = 1$ day or a few days, the orbital effect is very small.
We define:

$$B_{orb} = 2\left(\frac{\omega - \omega_0}{\omega_r} + \frac{m}{2} + \frac{n}{2}\frac{\omega_{orb}}{\omega_{rot}}\right)$$

$$\frac{\omega_r t}{2} = \xi \; : \; dt = \frac{2d\xi}{\omega_r} \quad (10)$$

$$\omega_r = R\omega_{orb}, \; (R \approx 365 \, days)$$

Next, we consider the integral:

$$I = \int_0^{2\pi} e^{-iB_{orb}\frac{\omega_r t}{2}} j_\ell\left(k\sin\left(\frac{\omega_r t}{2}\right)\right) dt \quad (11a)$$

Setting $\omega_r = R\omega_{orb}$ and defining

$$R\omega_{orb} t = y = s + 2\pi(j-1),$$

we have:

$$I = \int_0^{R2\pi} e^{-iB_{orb}\frac{y}{2}} j_\ell\left(k\sin\left(\frac{y}{2}\right)\right) \frac{dy}{R\omega_{orb}}, \; (R\omega_{orb} = \omega_r) \quad (11b)$$

The integral from 0 to $2\pi R$ can be split as a succession of sums from $[0,2\pi],[2\pi,4\pi]\ldots[(R-1)2\pi, R\,2\pi]$. Hence, the integral can be expressed as:

$$I = \sum_{j=1}^R \int_{(j-1)2\pi}^{j\cdot 2\pi} e^{-iB_{orb}\left(\frac{s+2\pi(j-1)}{2}\right)} \cdot j_\ell\left(k\sin\left(\frac{s}{2} + \frac{2\pi}{2}(j-1)\right)\right) ds \quad (12a)$$

Using the relation $j_\ell\left(k\sin\left(\frac{s}{2} + \pi(j-1)\right)\right) = e^{i\ell(j-1)\pi} j_\ell\left(k\sin\frac{s}{2}\right)$, the integrand with the summation becomes:

$$\sum_{j=1}^R e^{-\frac{i}{2}B_{orb} 2\pi(j-1)} j_\ell\left(k\sin\left(\pi\left(\frac{s}{2\pi} + j - 1\right)\right)\right) = Z[f(s, B_{orb})] \quad (12b)$$

where $Z[\;]$ is the *Zak-Gelfand Transform*[5] of:

$$f(s, B_{orb}) = e^{-i\frac{B_{orb}}{2}\cdot 2\pi(j-1)} j_\ell\left(k\sin\pi\left(\frac{s}{2\pi} + j - 1\right)\right) \quad (13)$$

This integral transform has been of use in the treatment of Bloch wave functions, for say, an electron in periodic potentials in condensed matter physics[6] and other applications of mathematical relevance.

Recalling

$$\frac{s}{2} = \xi = \frac{\omega_r t}{2},$$

we have

$$I = \sum_{j=1}^R \int_0^\pi e^{i\pi(j-1)\{\ell - B_{orb}\}} e^{-iB_{orb}\xi} j_\ell(k\sin\xi) d\xi. \quad (14)$$

The series related to the *Zak* transform can be summed:

$$\sum_{j=1}^{R} e^{i\pi(j-1)(\ell - B_{orb})} = \frac{1 - e^{i\pi(\ell - B_{orb})R}}{1 - e^{i\pi(\ell - B_{orb})}}, \quad (R = 365), \quad (15)$$

Here $(\ell - B_{orb}) \in \Re$ (real numbers) and $e^{i\pi R(\ell - B_{orb})} \in C$ (complex numbers).

The integral $I_1 = \int_0^\pi e^{-iB_{orb}\xi} j_\ell(k\,Sin\,\xi)d\xi$ can be evaluated exactly as well as numerically. The analytic expression is given below:

$$I_1 \equiv 2e^{-iB_{orb}\frac{\pi}{2}} \int_0^{\frac{\pi}{2}} Cos(B_{orb}\xi) j_\ell(k\,Cos\,\xi)d\xi$$

$$= 2e^{-iB_{orb}\frac{\pi}{2}} \cdot \sqrt{\frac{\pi}{2}} \int_0^1 Cos(B_{orb}\,Cos^{-1}x) \frac{J_{\ell+\frac{1}{2}}(kx)}{\sqrt{kx}} \frac{dx}{\sqrt{1-x^2}} \quad (16)$$

$$= 2e^{-iB_{orb}\frac{\pi}{2}} \cdot \pi \sqrt{\frac{\pi}{2}} k^{\ell+\frac{1}{2}} \cdot \Gamma\left[\ell+1, \ell+\frac{3}{2}, \frac{\ell + B_{orb} + 2}{2}, \frac{\ell - B_{orb} + 2}{2}\right] \cdot {}_1F_3\left(\ell+1; \ell+\frac{3}{2}, \frac{\ell + B_{orb} + 2}{2}, \frac{\ell - B_{orb} + 2}{2}; \frac{-k^2}{16}\right)$$

The $\Gamma[\ ]$ denotes one gamma function in the numerator, and three in the denominator.
The ${}_1F_3$ denotes the Generalized Hypergeometric Function with one parameter in the numerator, (here $\ell + 1$), and three in the denominator. The variable is $\frac{-k^2}{16}$.

$$\text{Hence: } \Sigma S_{n\ell m}(\omega_0, \omega_r, \omega_{orb}, d, t, A, k, \theta, \phi) = \begin{bmatrix} \sum_{n=-\infty}^{\infty} \sum_{\ell=0}^{\infty} \sum_{m=-\ell}^{\ell} 4\pi i^\ell Y_{\ell m}(\theta,\phi) N_{\ell m} P_\ell^m(Cos\,\alpha) \\ \cdot \sqrt{\frac{\pi}{2}} e^{-i\frac{2\pi f_0 A}{c} Sin\,\theta\,Cos\,\phi} \cdot i^n e^{-in\phi} J_n\left(\frac{2\pi f_0 A\,Sin\,\theta}{c}\right) \\ \cdot \left\{\frac{1 - e^{i\pi(\ell - B_{orb})R}}{1 - e^{i\pi(\ell - B_{orb})}}\right\} \cdot 2e^{-iB_{orb}\frac{\pi}{2}} \cdot \pi \cdot k^{\ell+\frac{1}{2}} \frac{1}{2^{2\ell+1}} \\ \cdot \frac{\Gamma(\ell+1)}{\Gamma\left(\ell + \frac{3}{2}\right) \Gamma\left(\frac{\ell + B_{orb} + 2}{2}\right) \Gamma\left(\frac{\ell - B_{orb} + 2}{2}\right)} \\ \cdot {}_1F_3\left(\ell+1; \ell+\frac{3}{2}, \frac{\ell + B_{orb} + 2}{2}, \frac{\ell - B_{orb} + 2}{2}; \frac{-k^2}{16}\right) \end{bmatrix} \quad (17)$$

$|\Sigma S_{n\ell m}|^2$ can also be directly calculated.

## 5  Discussion

We ran Maple programs using the analytical formulation with the following parameters: $\Omega = \frac{\omega - \omega_0}{\omega_r} = 0.9$, $m = 0$, $\ell$ *from* 1 *to* 301 *in steps of* 1. Plots for these calculations are shown in Figure 1. These plots show that a non-zero signal value was produced for $\ell$ values between 100 and 170, indicating the remarkable cut-off properties of the Bessel Index Theorem. Other runs were done for $\Omega = 20.4$ *and* $25.9$ *with* $m = 30$ which showed that the signal intensity decreased substantially as $\ell$ was increased. For low values of *k*, the computer time should be substantially reduced. We note that no signal was present when $B > \ell + 2$. Here $B = \left(2\left(\frac{\omega - \omega_0}{\omega_r} + \frac{m}{2} + \frac{n}{2}\frac{\omega_{orb}}{\omega_r}\right)\right)$. Maple programs using numerical integration (double exponential quadrature) for a wide range of parameters to enable a better comparison with the analytical work are planned in forthcoming work.

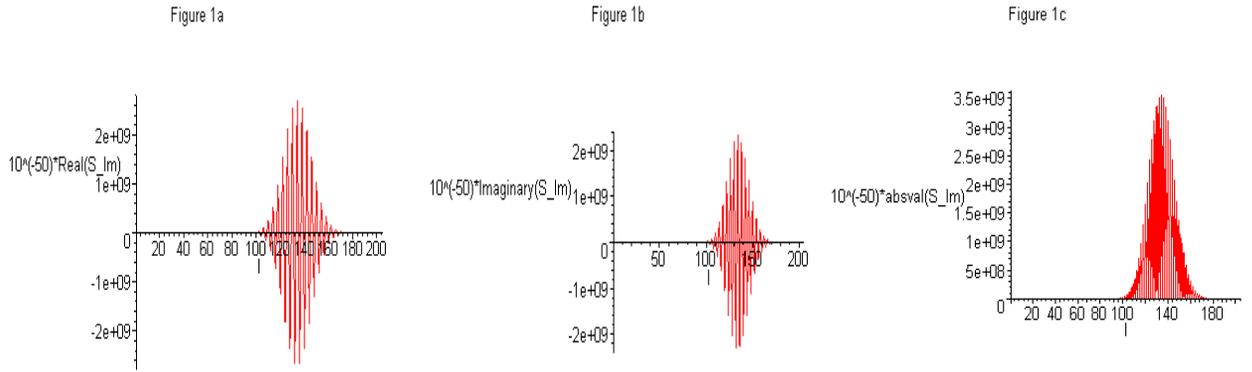

Figure 1a   Figure 1b   Figure 1c

## 6   Conclusion

We have studied the frequency modulation of the pulsar signal by two different methods. A closed form of the Fourier transform of the frequency modulated GW pulsar signal due to rotational and orbital motion of the Earth has been obtained. The work with the inclusion of the orbital corrections results in an interesting series related to the Zak-Gelfand integral transform. The Earth's position in longitude and radius vector will be affected by the perturbation due to Jupiter. The Jovian perturbation can increase or decrease the Earth's solar distance by 0.0016 %. Further details of this analysis warrant study. Such results would be useful for schemes like the stepping around the sky method[3] and for differential geometric methods that allow the setup of search templates in the relevant parameter space.[7] This analysis should also be applicable to the Doppler modulation of the GW signal caused by LISA's orbit around the sun. The physics and astrophysics that can be explored by GW observations in all wavebands will usher in a new trend in the field of gravitational wave phenomenology.

## 7   Acknowledgments

We thank Professor Tom Prince of Caltech and Professor Curtis Wilson of St. John's College, Annapolis, Maryland for their helpful comments. We thank Bob Byers of Stanford for an inspiring poster session at the Amaldi conference. We also thank Dave Fielder and Myra Fang for superb design and assistance with the poster. Finally, SRV thanks the Amaldi conference organizers for partial financial support and the Natural Sciences Engineering Research Council Canada (NSERC) for an Operating Research Grant.

**References**

1. A. Abramovici et al, Science, **256**, 325 (1992).
2. K. Jotania, S. Valluri, and S.V. Dhurandhar, Astron & Astrophys **306**, 317 (1996).
3. B.F. Schutz, in *The Detection of Gravitational Waves,* ed. D.G. Blair (Cambridge University Press, 1991).
4. B.H. Bransden & C.J. Joachain, *Introduction to Quantum Mechanics,* Second Edition, Prentice Hall 2000, and Los Alamos archives paper astro-ph/12002 v1 1 Dec. 2000 "A Study of the Gravitational Wave Form From Pulsars", Valluri et. al.
5. Ahmed I. Zayed, *Handbook of Function and Generalized Function Transformations,* CRC Press 1996.
6. J. Zak, *Dynamics of electrons in solids in external fields,* Phys. Rev., **168**, 686-695 (1968), *Finite translation in solid state physics,* Phys. Rev. Lett., **19**, 1385-1389 (1967).
7. B.S. Sathyaprakash, Filtering gravitational waves from Supermassive Black Hole Binaries, *LISA Conference, CALTECH* (1998).